\documentclass[aip, jap, 10pt, twocolumn, reprint, superscriptaddress, amsmath, amssymb]{revtex4-2} 

\usepackage{graphicx}
\usepackage{dcolumn}
\usepackage{bm}
\usepackage{ulem}
\usepackage{hyperref}
\usepackage[usenames]{xcolor}

\usepackage{multirow}
\usepackage{threeparttable}

\begin{document}


\title{Intrinsic Carrier Losses in Tellurium Due to Radiative and Auger Recombinations}

\author{J{\"o}rg Hader}
\email[Author to whom correspondence should be addressed: ]{jhader@acms.arizona.edu}
\affiliation{%
Wyant College of Optical Sciences, University of Arizona, 1630 E. University Blvd., Tucson, Arizona 85721, USA
}%

\author{Sven C. Liebscher}%
\affiliation{ 
Department of Physics and Material Sciences Center, Philipps-University Marburg, Renthof 5, 35032 Marburg, Germany
}%

\author{Jerome V. Moloney}
\affiliation{%
Wyant College of Optical Sciences, University of Arizona, 1630 E. University Blvd., Tucson, Arizona 85721, USA
}%

\author{Stephan W. Koch}
\affiliation{ 
Department of Physics and Material Sciences Center, Philipps-University Marburg, Renthof 5, 35032 Marburg, Germany
}%

\date{\today}

\begin{abstract}
Fully microscopic many-body models based on inputs from first principle density functional theory are used to calculate the carrier losses due to radiative-  and Auger-recombinations in bulk tellurium.  It is shown that Auger processes dominate the losses for carrier densities in the range typical for applications as  lasers.  The Auger loss depends crucially on the energetic position of the $H_6$ valence bands. At cryogenic  temperatures of 50$\,$K (100$\,$K) the Auger coefficient, $C$, varies by about six (three) orders of magnitude within the range of published distances between these bands and the valence bandedge. Values for $C$ at the high and low end of these ranges are found if the distance is smaller or larger than the bandgap, respectively. At room temperature the sensitivity is reduced to about a factor of four with $C$ values ranging between $0.4$ and $1.6\times 10^{-27}$cm$^6$s$^{-1}$. Here, radiative losses dominate for carrier densities up to  about $10^{16}/$cm$^3$ with a loss coefficient $B\approx 10^{-11}$cm$^3$s$^{-1}$. The radiative losses are about two to three times lower than in typical bulk III-V materials for comparable  wavelengths.   
\end{abstract}

\maketitle

Triagonal tellurium (t-Te) combines the simplicity of a single elemental material with a wide variety of technically interesting properties. Its distinctive  structure made of weakly interacting helical chains of atoms and triagonal symmetry perpendicular to the axis of the chains, $c$,  lead to a strong anisotropy of its characteristics like an exceptionally strong  optical nonlinearity\cite{nonlin65} or an extraordinarily high piezoelectric coefficient\cite{png13}. Recently, the interest in tellurium has intensified as it was shown that it transforms into a topological insulator under pressure\cite{topo_ins14} and it has become possible to synthesize it in the two-dimensional form tellurene and as one-dimensional wires\cite{fewlayerte18,tellurene22}.

t-Te is a semiconductor with a bandgap in the mid-infrared wavelength range near $3.7-3.8\,\mu m$. Reaching these wavelengths using conventional III-V materials requires complex heterostructures like quantum cascade systems, or involve complex mixtures of multiple binary materials, like quinternary AlInGaAsSb, which are difficult to grow in a controlled manner\cite{midir19}. The size of this bandgap combined with a high conductivity make the material ideal for applications as a (chiral) thermoelectric material\cite{te_bands_prb14,thermoelectric16}. 

While t-Te has an indirect gap due to a camel-back structure of the highest valence band, the dip in the valence band is only of the order of 1$\,$meV\cite{te_bands_prb14,te_indir_exp70} and the material behaves like a direct semiconductor under common conditions. This makes t-Te a candidate for optoelectronic applications like photo-detectors, solar cells or light emitters\cite{te_review22}. To be successful for these applications requires good crystal quality, strong optical coupling and low carrier losses. The atomic simplicity of t-Te enables high crystal quality and measured\cite{exp_abs_te69} and calculated\cite{prb_te21} absorption spectra have confirmed strong optical coupling at the bandgap. The latter was found very similar to that  in bulk III-V materials with a similar bandgap, like InAs\cite{exp_abs_inas16} or dilute InAsBi\cite{apl_inasbi18}.
  
The high crystal quality of t-Te allows for low defect related carrier losses\cite{te_recomb_jpc19}. Carrier losses in t-Te due to radiative recombinations have recently been measured in Refs.\cite{te_recomb_jpc19} and Ref.\cite{te_recomb_prb19}. The works determined  recombination coefficients, $B$, of 1.1$\times$10$^{-8}$cm$^3/$s and 1.9$\times$10$^{-9}$cm$^{3}/$s, respectively. Both values are significantly larger than the commonly accepted values in direct gap bulk III-V materials like GaAs\cite{b_gaas81,b_gaas93,b_gaas05,b_gaas19}, GaSb\cite{semicon_param00}, InP\cite{semicon_param00}, InSb\cite{semicon_param00}, or InAs\cite{semicon_param00}. For these, $B$ has been determined to be in the range of 0.2-2$\times$10$^{-10}$cm$^3/$s. Generally, the radiative recombination due to spontaneous emission should scale like the near bandgap absorption since in good approximation the two can be directly related through the Kubo-Martin-Schwinger (KMS) relation\cite{kms99}. Since the absorption in t-Te is of similar strength as in these other materials, the radiative lifetimes should not be expected to be significantly different either.

The large magnitude and spread of the values for $B$ in Refs.\cite{te_recomb_jpc19,te_recomb_prb19} are likely due to the fact  that the material is inherently p-doped. The doping arises due to the fact that the chiral chains are finite and unsaturated bonds occur at their ends. The exact level of doping is generally not known and sample-dependent. The intrinsic radiative lifetime that assumes equal electron and hole carrier densities can be vastly different from the minority carrier lifetime in doped samples and the latter strongly depends on the dopant level\cite{yu_cardona10}.

Carrier losses due to Auger recombinations are known to increase dramatically with decreasing bandgap\cite{bewley08}. For materials with bandgaps similar to t-Te,  like InAs or InSb, Auger rates have been found to be about three orders of magnitude higher than in materials with a bandgap around one micron, like GaAs (Auger coefficients, $C$ around 10$^{-27}$cm$^6$/s versus 10$^{-30}$cm$^6$/s\cite{apl_inasbi18,landoldt02,landoldt09}). This makes Auger loss the limiting factor for mid-IR materials in applications as optoelectronic devices. To the best of our knowledge, Auger rates have neither been calculated nor measured for t-Te. Generally, measuring the losses experimentally is difficult since only the total non-radiative recombination rate is accessible, but the Auger contribution cannot be measured individually. 

Knowing the intrinsic recombination rates allows to judge the prospects of t-Te for applications. Here we use fully microscopic many-body models based on inputs from first principle density functional theory (DFT) in order to calculate the intrinsic carrier losses due to radiative- and Auger-recombination. These models have been shown to yield excellent agreement with the experiment for a wide variety of materials with wavelengths ranging from the mid-IR to the ultra-violet\cite{nlcstr}. The {\it a-priori} nature of the modeling approach eliminates fit parameters that would otherwise be required to be extracted from experiments, like linewidth broadenings or dephasing times. Elimination of such parameters, combined with the demonstrated accuracy of the results makes this approach quantitatively predictive.

The carrier lifetimes due to spontaneous emission processes, also known as bimolecular recombination lifetimes or, short, as radiative carrier lifetimes, are calculated using the semiconductor luminescence equations (SLE)\cite{sle99}. These are the equations of motion for microscopic photon-assisted polarizations.  They explicitly contain carrier-carrier and carrier-phonon scatterings that lead to the dephasing of the polarizations and influence  photoluminescence (PL) lineshapes, spectral positions and amplitudes. The corresponding terms also enter the SLE as part of higher order correlations that are sources for the photon-assisted polarizations. These source terms are absent in the equations of motion for the optical polarizations that describe absorption/gain spectra. This absence leads to an error when using the KMS relation to derive the PL from the absorption that can easily be on the order of a factor of two or more. The PL spectrum is obtained from the total photon assisted polarization via Fourier transformation and the radiative lifetime is obtained by integrating over the spectrum\cite{trad_frompl99}.

Like the scattering terms involved in the SLE, Auger losses are calculated in second-Born and Markov approximation by solving quantum-Boltzmann type scattering equations\cite{auger_theo05}. The equations that we solve for the Auger losses are basically the same as derived  decades earlier\cite{auger_theo58}. However, unlike earlier attempts to solve these equations, we do not use any uncontrolled approximations like using Boltzmann instead of Fermi distributions, approximating the Coulomb coupling matrix elements or considering only occupations near zero momentum or at 0$\,$K. These approximations lead to errors that can easily exceed one order of magnitude and are to a large degree the reason why values for Auger losses in the literature vary for many materials by such large margins. As mentioned, our calculations have been tested against the experiment for a wide variety of materials, temperature- and density-ranges and have shown uncertainties in the few ten-percent range\cite{nlcstr,sle99}.

We assume equal electron and hole carrier densities $N$ and all Auger calculations are for direct inter-band processes which are dominant in materials for this wavelength range\cite{aug_inas19}. The bimolecular (radiative) recombination coefficients, $B$, and Auger coefficients, $C$ are derived from the corresponding  recombination times, $\tau_B$ and $\tau_C$, through:
\begin{equation}
\label{eq1}
B = \frac{1}{\tau_B\,\,N},\qquad C = \frac{1}{\tau_c\,\,N^2}.
\end{equation}

The band energies and wavefunctions that enter the SLE and Auger equations are calculated using the Vienna Ab Initio Simulation Package\cite{kresse1, kresse2, kresse3, kresse4} (VASP)  with the Projector-Augmented Wave (PAW) method\cite{kresse5, blochl1994}.
The structure was relaxed using the Generalized Gradient Approximation (GGA) by Perdew, Burke and Ernzerhof (PBE)\cite{perdew1996} for the exchange-correlation energy. The PAW pseudopotential was modified according to the shLDA-1/2 method as proposed In Ref.\cite{xue2018}. More details of the DFT calculation can be found in Ref.\cite{prb_te21}.

\begin{figure}[htp]
\begin{center}
\includegraphics[width=1.0\linewidth]{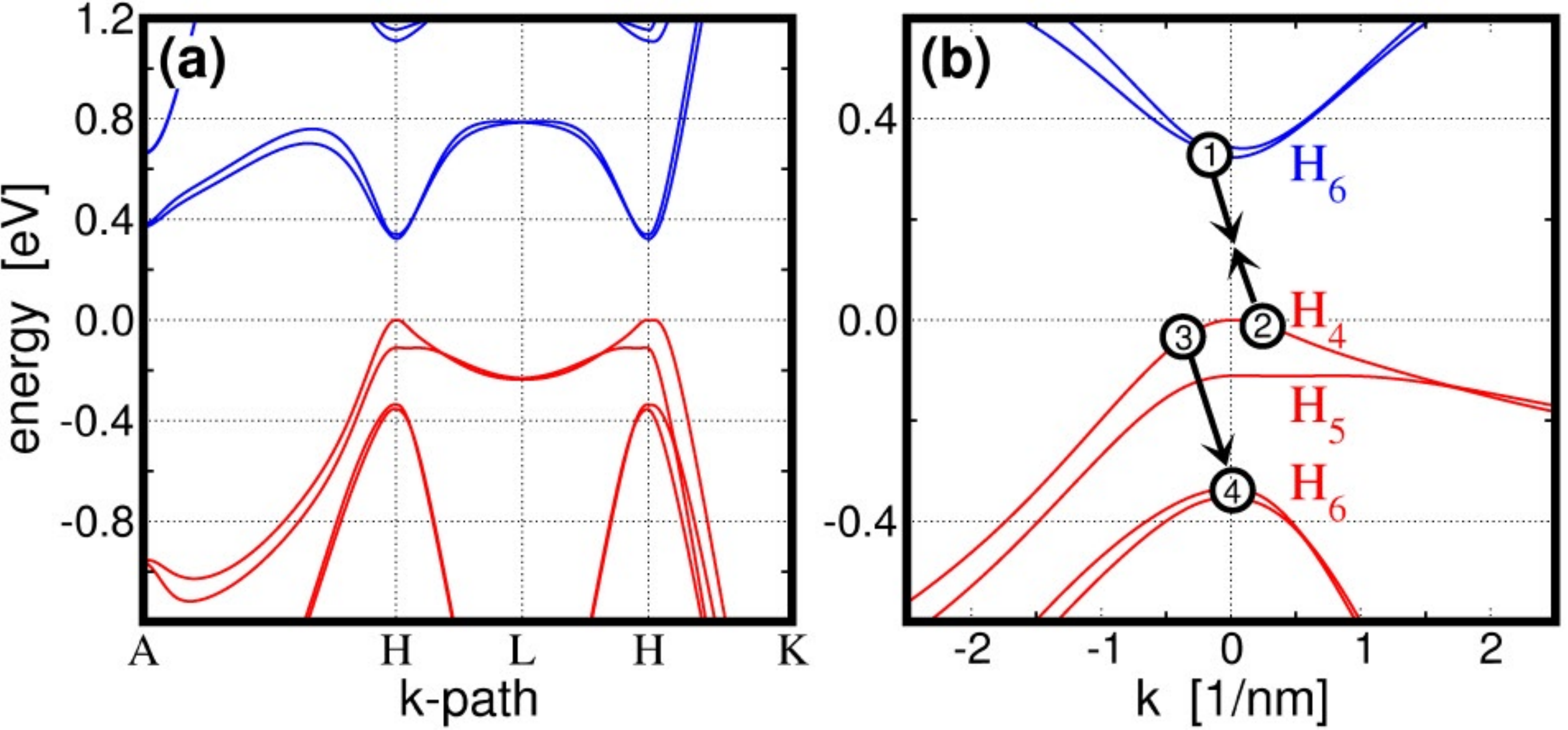}
\end{center}
\caption{(a): Lowest four electron (blue) and highest hole (red) bands along some high symmetry directions of the Brillouin zone of t-Te as calculated with DFT. (b): Zoom into the region around H along the A--H--L path with the schematic of an Auger recombination of an electron (1) with a hole (2) where the excess energy is being given to another hole (3) which is excited into a state (4) deep in the valence band.}
\label{fig1}
\end{figure}

Fig.\ref{fig1} shows the results of the bandstructure calculation for t-Te along some high symmetry directions of the Brillouin zone (BZ). Like Ref.\cite{te_bands_prb14}, we find a small camel-back shape at the top of the valence band edge with a global maximum slightly away from the $H$-point and about 1$\,$meV above the energy at $H$. At 0$\,$K the bandgap at the $H$-point is $E_0=\,$333$\,$meV and the splitting between the highest two valence bands $E_{H_4-H_5}=\,$111$\,$meV. The splitting between the highest and third highest valence bands $E_{H_4-H_6}=\,$337$\,$meV. The values for the gap and $E_{H_4-H_5}$ agree very well with experimentally measured values\cite{te_gapexp59,te_gapexp77,te_gapexp22} and other calculations\cite{topo_ins14,te_bands_prb14,te_gaptheo18}. However, there is some uncertainty in the literature for $E_{H_4-H_6}$. While bandstructure shown in Ref.\cite{te_bands_prb14} agrees with our value, Ref.\cite{te_gapexp22} extracted a value of 427$\,$meV from experiments and Ref.\cite{te_gaptheo18} calculates a splitting of about 415$\,$meV. The error margin of the measured value in Ref.\cite{te_gapexp22} is not clear, but could be on the order of several tens of meV. 

As we will show below, the exact value for $E_{H_4-H_6}$ has a significant impact on the Auger losses. At the low end of the range of published values the splitting is almost exactly equal to the bandgap. This is an ideal situation for Auger transitions since it allows holes to be excited from the $H_4$ to the $H_6$ band with minimal momentum transfer as indicated in the schematic in Fig.\ref{fig1} (b). On the other hand, for the larger values, the band separation is larger than the bandgap and carriers near the gap are not able to reach the $H_6$ bands. Here, the only possible transitions are within the $H_4$ and $H_5$ bands or within the $H_6$ electron band. These require large momentum transfers to reach states one bandgap away from the bandedge and are therefore unfavorable since the Coulomb coupling decreases with increasing momentum transfer. 

To compare results for the two different splittings we apply for the larger splitting case a momentum-independent shift to the $H_6$ hole bands from our calculation. All other bands and all wavefunctions are kept the same.

The DFT calculations are performed for a temperature of 0$\,$K. We include the temperature dependence of the bandstructure by applying a shift to the conduction bands according to the formula for the bandgap as derived from the experiment in Rev.\cite{te_gapexp22}. This leads to a gap that has a maximum of 335$\,$meV around 75$\,$K, is about 333$\,meV$ at 0$\,K$ and 327$\,$meV at 300$\,K$. For temperatures above 300$\,$K the bandgap shrinks by about 0.052$\,$meV/K.

\begin{figure}[htp]
\begin{center}
\includegraphics[width=1.0\linewidth]{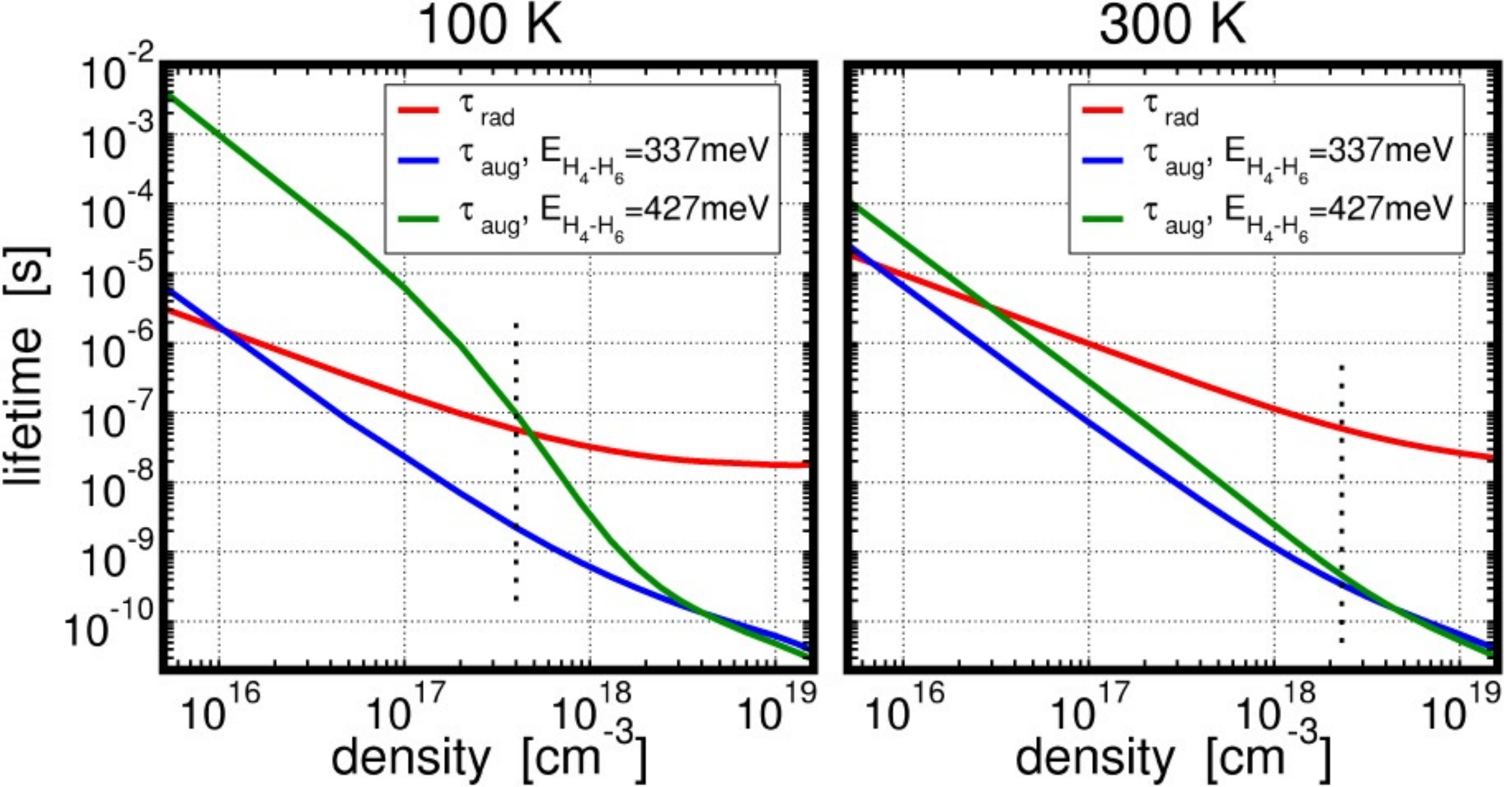}
\end{center}
\caption{Calculated carrier lifetimes due to radiative (red) and Auger recombinations as function of the carrier density at 100$\,$K (left) and $300\,$K (right). Blue (green): Auger lifetimes for the case that the hole splitting $E_{H_4-H_6}$ is similar to  (337$\,$meV)  (larger than (427$\,$meV)) the bandgap. Dotted  lines mark the approximate carrier density at which the absorption at the bandgap is bleached and optical gain emerges.}
\label{fig2}
\end{figure}

Fig.\ref{fig2} shows the calculated carrier lifetimes due to radiative and Auger recombinations as function of the carrier density. Auger losses are shown for two cases of the hole splitting $E_{H_4-H_6}$, the case where it is almost resonant with the bandgap (337$\,meV$) and the case where it is about 95$\,$meV larger than the bandgap (427$\,$meV). At 300$\,$K Auger losses dominate already at densities on the order of 10$^{17}$/cm$^3$ which is more than one order of magnitude below the transparency density at that temperature. At 300$\,$K the Auger losses at low densities are about four times smaller if $E_{H_4-H_6}$ is larger than the gap rather than when $E_{H_4-H_6}$ and the gap are almost equal. At 100$\,$K this difference is close to three orders of magnitude. 

At the higher temperature the larger high energy tail of the Fermi distributions leads a larger percentage of holes at energies far enough below the bandedge to  reach the $H_6$ bands even for the larger splitting. This increases the Auger loss for the large splitting case and makes it more similar to the small splitting case. Similarly, the sensitivity to the splitting decreases with increasing density since an increasing fraction of carriers occupies states sufficiently far away from the gap to reach the $H_6$ bands.  

In the limit of very high densities the Auger losses for the larger splitting start to exceed somewhat those for the smaller splitting. This is due to the fact that for the larger splitting at these densities more carriers can reach  final states with less momentum transfer and larger Auger coupling.

The radiative lifetimes are virtually identical for both values for $E_{H_4-H_6}$. This is due to the fact that for both cases of the splitting virtually no carriers occupy the $H_6$ hole bands even at the highest temperatures and densities considered here. Non-zero occupations occur only for the $H_4$ and $H_5$ hole bands and these occupations are not affected by the position of the $H_6$ band. Since the spontaneous emission only involves occupied states it is unaffected by the splitting and so are the resulting lifetimes.

\begin{figure}[htp]
\begin{center}
\includegraphics[width=1.0\linewidth]{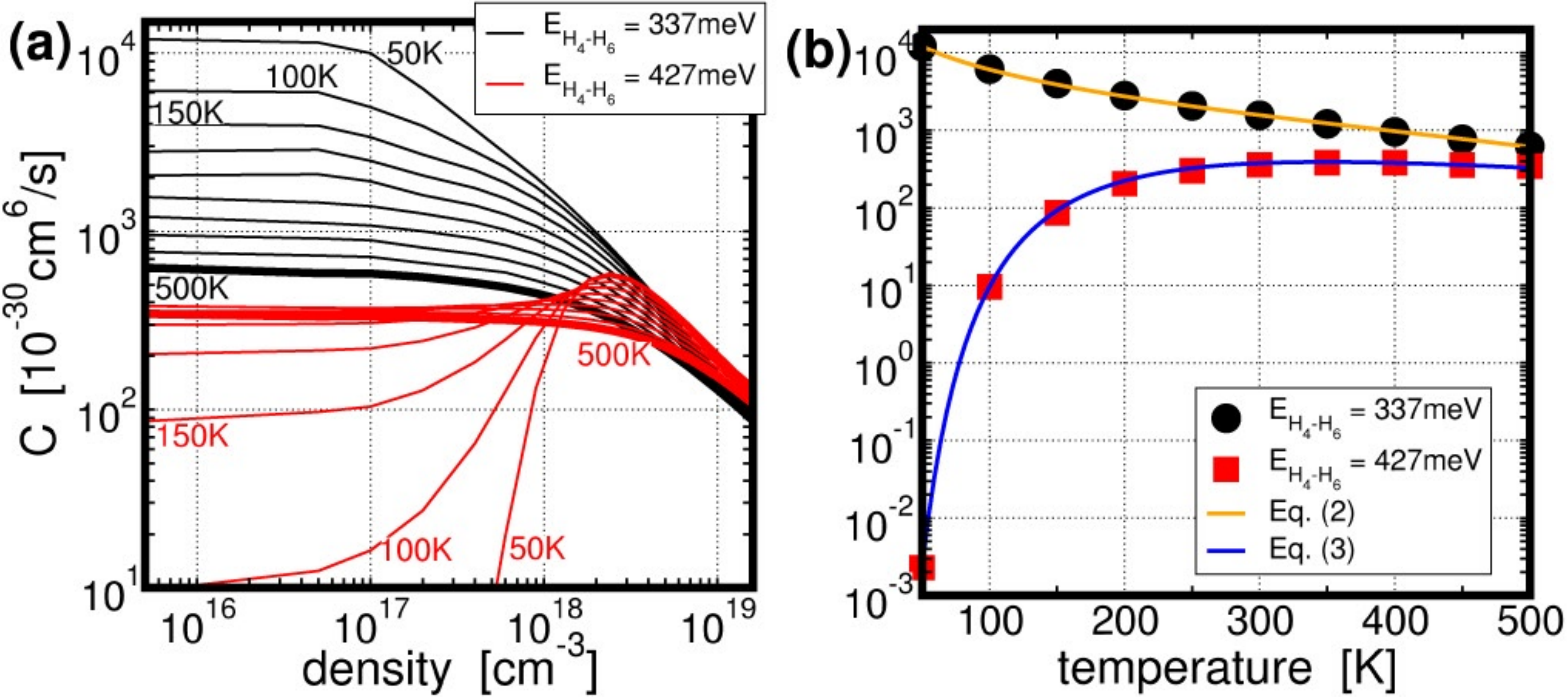}
\end{center}
\caption{(a) Auger coefficient as function of the carrier density for temperatures between 50$\,$K and 500$\,$K in 50$\,$K intervals. The result for 500$\,$K is marked with a bold line. Black (Red): for a hole splitting $E_{H_4-H_6}$=337$\,meV$ (427$\,$meV). (b) Auger coefficients in the low density limit as function of the temperature. Circles (Squares): calculation results for $E_{H_4-H_6}$=337$\,meV$ (427$\,$meV). Orange (blue) Line: fit according to Eq. (\ref{eq2}) (Eq. (\ref{eq3})). }
\label{fig4}
\end{figure}

Fig.\ref{fig4} shows the Auger coefficients as function of the density for various temperatures. In the limit of low densities ($N\lessapprox 10^{17}/cm^3$) one finds the traditional quadratic density dependence of the lifetimes and, thus, density independent coefficients $C$. In the high density limit ($N\gtrapprox 5\times 10^{18}/cm^3$) the density dependence is reduced due to phase space filling\cite{apl_psf05}. At low densities the occupation probabilities increase approximately linearly with the density leading to the quadratic density dependence of the lifetimes. At high densities Pauli-blocking limits a further increase of the occupations and additional carrier have to be filled instead into states at higher energies. Here, this leads to a reduction of the density dependence of the Auger lifetimes from quadratic to about linear and, correspondingly, a nearly linear decrease of $C(N)$. 

The values found here for $C$ for the case where $E_{H_4-H_6}$ is larger than the bandgap are very similar to those in InAs. Like t-Te, InAs has a bandgap in the 3-4$\,\mu m$ range (356$\,$meV at 300$\,K$). Its spin-orbit hole splitting is larger than the bandgap and of similar magnitude as the $H_4-H_6$-splitting here (410$\,$meV). Its Auger coefficient at room temperature has been found to be around 10$^{-27}$cm$^6$/s\cite{apl_inasbi18}.

Fig.\ref{fig4} (b) shows the temperature dependence of the Auger coefficient in the low density limit. For the case where $E_{H_4-H_6}$ is about equal with the gap $C$ decreases with increasing temperature. This is mostly due to the fact that the occupation probabilities near the bandgap decrease with increasing temperature. The dependence can be fitted using the expression: 
\begin{equation}
\label{eq2}
C(T) = \frac{7.4\times 10^{-25} - 8.6\times 10^{-28}\,T}{T+8}\,\, [cm^6/s],
\end{equation}
where $T$ is the temperature in Kelvin.

The low density $C(T)$ increases dramatically at low temperatures if $E_{H_4-H_6}$ is larger than the bandgap. In the cryogenic limit, no carriers are at energies close enough to reach the $H_6$ band and transitions within the near-gap bands are unfavorable since they require large momentum transfer. This leads to very low Auger rates. With rising temperature an increasing fraction of holes occupies states close enough to the $H_6$-band to allow for  Auger transitions to it. This leads to a strong increase of $C(T)$ for temperatures up to about room temperature. For even higher temperatures the trend seen for the case where the splitting is smaller than the gap starts to set in and $C(T)$ starts to decrease. The behavior for low temperatures can be described by an activation energy law with an energy similar to the difference between the splitting and the gap. The decrease at higher temperatures leads to a modification of the dependence that can be modeled by an additional $1/T^3$-scaling:
\begin{equation}
\label{eq3}
C(T) =  3.3\time 10^{-19} \times exp\left(\frac{-E_a}{k_B\,T}\right)T^{-3}    \,\, [cm^6/s],
\end{equation}
where $k_B$ is the Boltzmann constant and $E_a$=0.09$\,$eV.

\begin{figure}[htp]
\begin{center}
\includegraphics[width=1.0\linewidth]{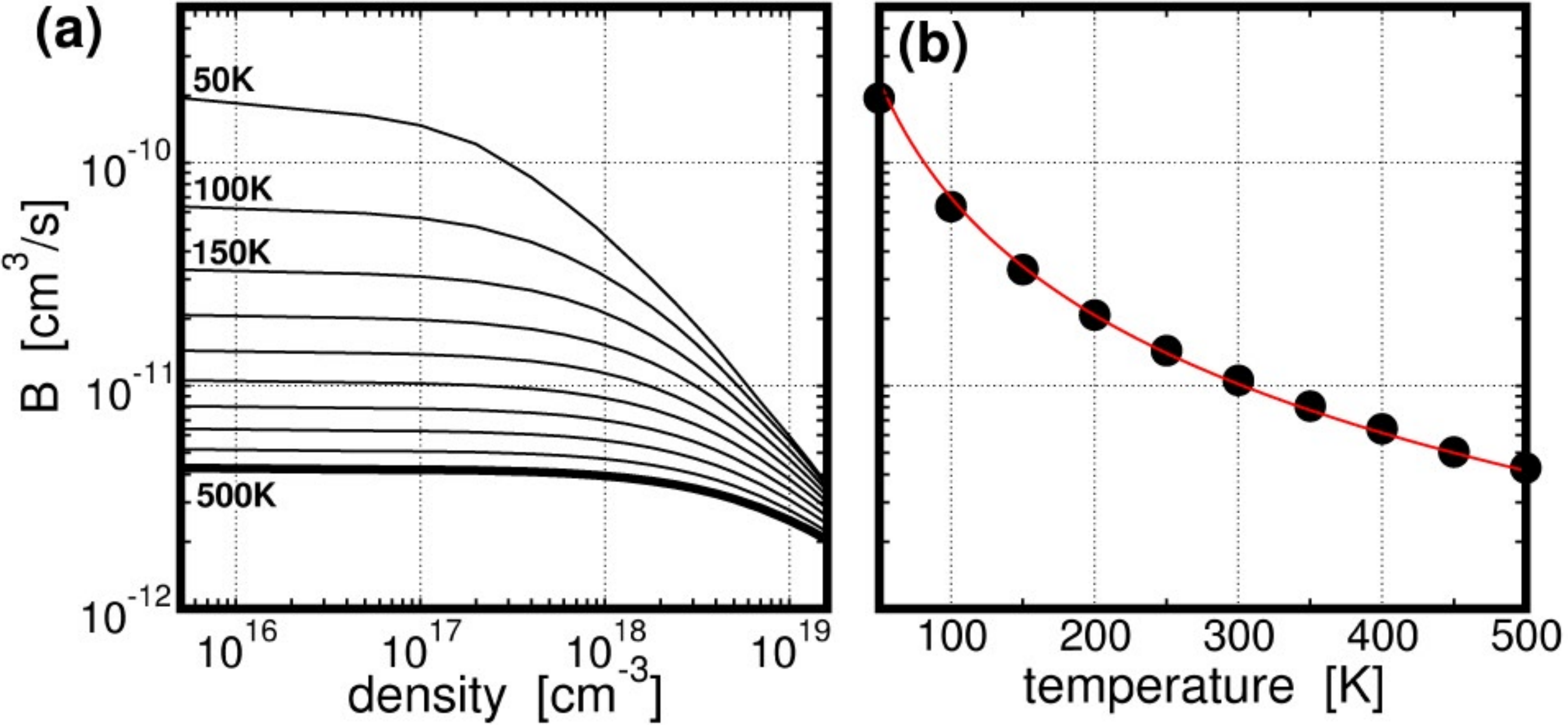}
\end{center}
\caption{(a) Bimolecular recombination coefficient, $B$ as function of the carrier density for temperatures between 50$\,$K and 500$\,$K in 50$\,$K intervals. (b) Coefficients $B$ in the low density limit as function of the temperature. Circles show calculation results. The red line is a fit according to Eq. (\ref{eq4}).}
\label{fig5}
\end{figure}

Fig.\ref{fig5} shows the calculated bimolecular recombination coefficient $B$ as function of the carrier density and for various temperatures. As mentioned above, the $H_4-H_6$ hole splitting has virtually no influence on the radiative recombination rate. Within the parameter range investigated here we find the differences between results for the two splittings on the order of 2$\times$10$^{-4}$ and indistinguishable on the scale of Fig.\ref{fig5}. 

In the limit of low carrier densities we find $B$(77$\,$K)$\approx\,$1.1$\times$10$^{-10}$cm$^3$/s and $B$(300$\,$K)$\approx\,$1.1$\times$10$^{-11}$cm$^3$/s. The temperature dependence of the low density value of $B$ as shown in Fig.\ref{fig5} can be fitted using:
\begin{equation}
\label{eq4}
B(T) =  2.2\times 10^{-7} \, T^{-7/4}    \,\, [cm^3/s],
\end{equation}

As for Auger recombination, the bimolecular recombination rates found here are similar to those reported for III-V bulk materials at similar wavelengths. Ref.\cite{apl_inasb12} measured a coefficient $B$ of 3$\times$10$^{-10}$cm$^3$/s at 77$\,$K in InAsSb. Using SimuLase$^{TM}$ software\cite{simulase}, which implements the same level of microscopic many-body physics as used here, we calculate for bulk InAs $B$(77$\,$K)$\approx$2.0$\times$10$^{-10}$cm$^3$/s and $B$(300$\,$K)$\approx$2.5$\times$10$^{-11}$cm$^3$/s. 

The radiative lifetimes in those III-V materials are about two to three times longer than what we find for t-Te. In part this is likely due to the fact that the optical coupling for polarization parallel to the c-axis is symmetry forbidden at the bandgap of t-Te. This effect which is reflected in the suppressed near-bandgap absorption of t-Te\cite{exp_abs_te69,prb_te21} also leads to a limited spontaneous emission into that direction\cite{prb_te21}.

Like the Auger coefficients, the bimolecular recombination coefficients decrease with density due to phase-space filling\cite{apl_psf05}. In the high density limit the coefficients tend toward a 1/$N$ density dependence.

In summary, we have used fully microscopic many-body models based on input from first-principle DFT calculations to study intrinsic carrier losses in bulk triagonal tellurium. Assuming that the splitting between the $H_4$ and $H_6$ valence bands is as in our calculations about equal to the bandgap we find Auger recombination losses of similar strength as in bulk III-V materials for similar wavelengths, like InAs. Auger coefficients of about  10$^{-27}$cm$^6$/s at 300$\,K$ make it unlikely that the material is viable for laser applications. However, the Auger losses could be dramatically lower if the splitting is larger than the bandgap as suggested in some other publications. Independent of the $H_4-H_6$ splitting, the radiative losses are about a factor three lower than in III-V materials for these wavelengths. This means that t-Te could yield better performance for low-density applications like photo detectors or solar cells.  

The authors thank the HRZ Marburg and CSC-Goethe-HLR Frankfurt for computational resources. The Tucson work was supported by the Air Force Office of Scientific Research under award numbers FA9550-19-1-0032 and FA9550-21-1-0463.

\section*{Data Availability}
The data that supports the findings of this study are available within the article.

\section*{ORCID iDs}

\noindent J{\"o}rg Hader  https://orcid.org/0000-0003-1760-3652\\
Sven C. Liebscher  https://orcid.org/0000-0003-3505-1521\\
Jerome V. Moloney  https://orcid.org/0000-0001-8866-0326\\
Stephan W. Koch  https://orcid.org/0000-0001-5473-0170

\end{document}